\documentclass{article}
\usepackage{spconf,amsmath,graphicx,adjustbox,booktabs,multirow}
\PassOptionsToPackage{hyphens}{url}\usepackage{hyperref}

\title{Wespeaker: A Research and Production oriented \\ Speaker Embedding Learning Toolkit}
\name{\begin{tabular}{c}
Hongji Wang$^{1,2,7}$, Chengdong Liang$^{3,4,7}$, Shuai Wang$^{1,7,*}$, Zhengyang Chen$^{1}$, Binbin Zhang$^{4,7}$, \\Xu Xiang$^{5}$, Yanlei Deng$^{6}$, Yanmin Qian$^{1}$\thanks{* Shuai Wang is the corresponding author.}
\end{tabular}
}
\address{$^{1}$X-LANCE Lab, Shanghai Jiao Tong University, Shanghai, China \\
$^{2}$Tencent Ethereal Audio Lab, Tencent Corporation, Shenzhen, China \\
$^{3}$School of Marine Science and Technology, Northwestern Polytechnical University, Xi'an, China \\
$^{4}$Horizon Robotics, Beijing, China,$^{5}$AISpeech Ltd, Suzhou, China,$^{6}$NVIDIA, Santa Clara, USA  \\
$^{7}$WeNet Open Source Community}

\begin{document}
\ninept
\maketitle
\begin{abstract}
Speaker modeling is essential for many related tasks, such as speaker recognition and speaker diarization. The dominant modeling approach is fixed-dimensional vector representation, i.e., speaker embedding. This paper introduces a research and production oriented speaker embedding learning toolkit, Wespeaker. Wespeaker contains the implementation of scalable data management, state-of-the-art speaker embedding models, loss functions, and scoring back-ends, with highly competitive results achieved by structured recipes which were adopted in the winning systems in several speaker verification challenges. The application to other downstream tasks such as speaker diarization is also exhibited in the related recipe. Moreover, CPU- and GPU-compatible deployment codes are integrated for production-oriented development. The toolkit is publicly available at \url{https://github.com/wenet-e2e/wespeaker}.
\end{abstract}
\section{Introduction}
Deep speaker embeddings~\cite{variani2014deep,snyder2018x,zeinali2019but,desplanques2020ecapa} are the de facto standard speaker identity representation for many related tasks. For speaker recognition, they are fed to scoring back-ends such as cosine similarity or probabilistic linear discriminant analysis (PLDA) for the following acceptance decision making. A similar application can be found in speaker diarization, where the scores obtained are used for further clustering. Besides the tasks that focus on speaker modeling, deep speaker embeddings are also utilized in other speech processing tasks, such as the speaker adaptation in speech recognition, speaker modeling in text-to-speech and voice conversion, etc.

\label{sec:intro}

Researchers in the intelligent speech processing community have been quite active in the open-source field. The early general speech processing toolkits, such as HTK~\cite{young2002htk} and Kaldi~\cite{povey2011kaldi}, had equipped many researchers and industrial productions before deep learning toolkits such as Pytorch~\cite{paszke2019pytorch} and Tensorflow~\cite{abadi2016tensorflow}, while the recently thriving Pytorch based SpeechBrain~\cite{ravanelli2021speechbrain}, Espnet~\cite{watanabe2018espnet}, are much more friendly for new researchers and enables fast prototyping. Unlike the mentioned general-purpose speech processing toolkits, Wenet~\cite{yao2021wenet, zhang2022wenet} focuses on end-to-end speech recognition and is designed to bridge the gap between the research and deployment.

% \textbf{Open-source toolkits}:
% \begin{itemize}
%     \item speech toolkits: kaldi/espnet/wenet/speech-brain 
%     \item speaker toolkits: asv-subtools, speech-brain 
%     \item relation to wenet
% \end{itemize}

Competitions such as the VoxSRC series~\cite{chung2019voxsrc,nagrani2020voxsrc,brown2022voxsrc} and CNSRC 2022 significantly promote the related dataset while accordingly inspiring researchers' creativity and engineering ability, leading to new SOTA results. However, there is usually a clear gap between the results reported in the research papers and competition system descriptions. The main reason might be that tricks and dedicated engineering efforts are usually neglected for the former. Another problem is that current speaker-related open-source implementations are research focused, without the support for potential migration for the production environment. 

To this end, we would like to design a speaker embedding learning toolkit that provides clean and well-structured codes for learning high-quality embeddings, with good portability to the production scenarios. Following Wenet, we propose an open-source Wespeaker which focuses on deep speaker embedding learning as another project of this ``we''-series. The key features/advantages of the Wespeaker toolkit are as follows,
\begin{itemize}
    \item \textbf{Competitive results}: Compared with other open-source implementations~\cite{ravanelli2021speechbrain, tong2021asv}, we achieve very competitive performance in all the recipes, including the VoxCeleb, CNCeleb, and VoxConverse. Many tricks used in the winning systems of the related competitions are re-implemented in Wespeaker to boost the system's performance. We hope Wespeaker can provide the researchers with a competitive starting point for their algorithm innovation.
    \item \textbf{Light-weight}: Wespeaker is designed specifically for deep speaker embedding learning with clean and simple codes. It is purely built upon PyTorch and its ecosystem, and has no dependencies on Kaldi~\cite{povey2011kaldi}.
    \item \textbf{Unified IO (UIO)}: A unified IO system similar to the one used in Wenet~\cite{zhang2022wenet} is introduced, providing a unified interface that can elastically support training with a few hours to millions of hours of data.
    \item \textbf{On-the-fly feature preparation}: Unlike the traditional feature preparation procedure, which performs utterance segmentation, data augmentation and feature extraction in an offline manner, Wespeaker performs all the above steps in an on-the-fly manner. Different augmentation methods, including signal-level ones such as noise corruption, reverberation, resampling, speed perturbation, and feature-level SpecAug~\cite{park2019specaugment}, are supported.
    %This online feature preparation procedure helps save disk usage while increasing the randomness and uncertainty of the training examples across different epochs.
    \item \textbf{Distributed training}: Wespeaker supports distributed training to speed up, where ``DistributedDataParallel'' in Pytorch is adopted for multi-node multi-GPU scalability.
    \item \textbf{Production ready}: All models in Wespeaker can be easily exported by torch Just In Time (JIT) or as the ONNX format, which can be easily adopted in the deployment environment. Sample deployment codes are also provided. 

\end{itemize}

% \vspace{-0.3cm}
\section{Wespeaker}
\label{sec:wespeaker}
\subsection{Deep speaker embedding learning}
For a standard deep speaker embedding learning system, the input is frame-level features (e.g., Fbank) and the expected output is segment-level embeddings. Such systems usually consist of several frame-level layers to process the input features, followed by a pooling layer to aggregate the encoded frame-level information into segment-level representations, and then several (commonly one or two) segment-level transform layers that map these representations to the correct speaker labels.
Moreover, a class-based or metric-based loss function is adopted to provide a speaker-discriminative supervision signal. 

\subsection{Overall structure}
\begin{figure}[!htb]
  \centering
  %\vspace{-0.1cm}
  \includegraphics[width=0.9\linewidth, trim=0 18 0 0, clip]{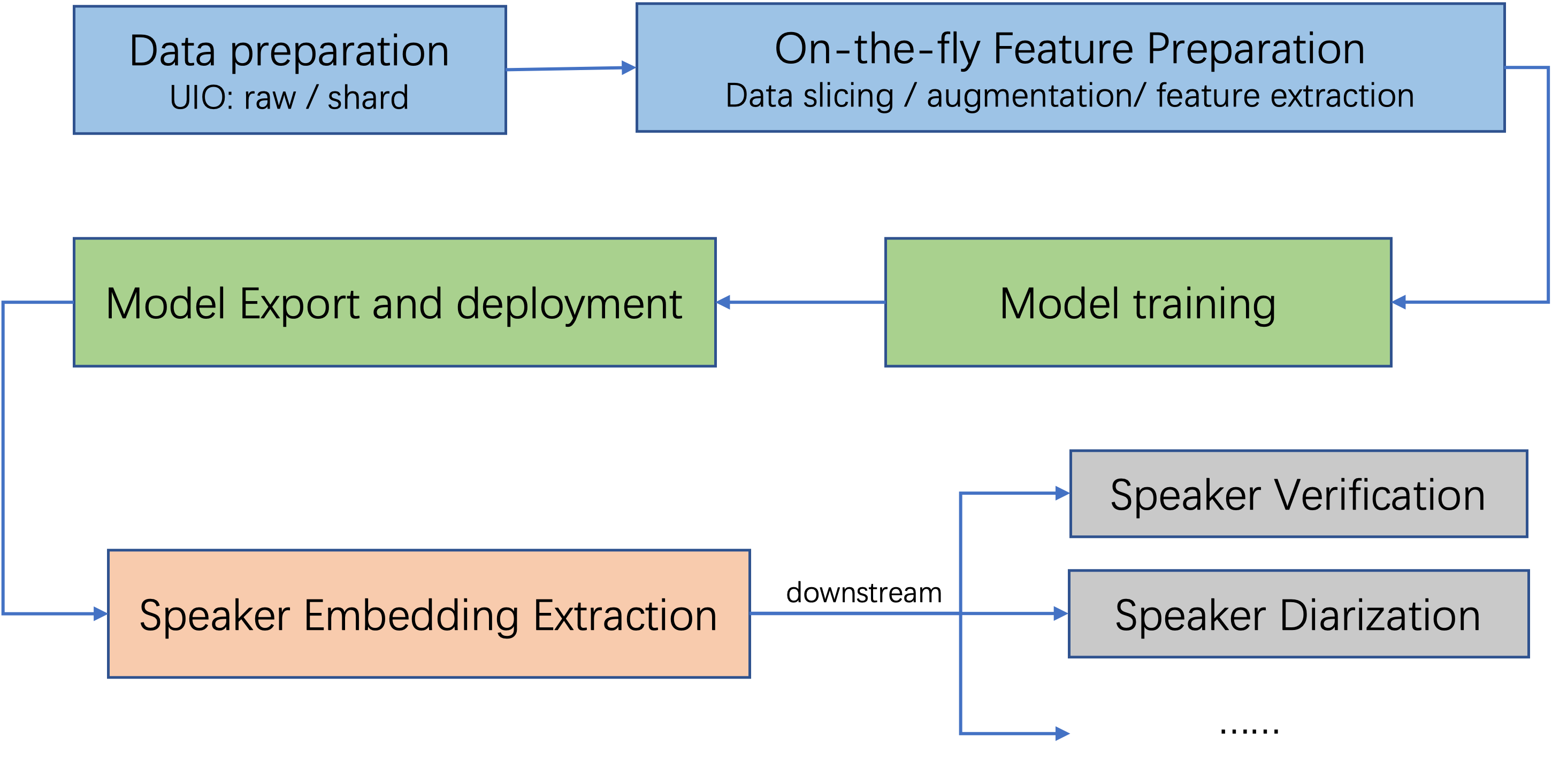}
  \caption{The overall structure of Wespeaker}
  \label{fig:wespeaker}
  \vspace{-0.5cm}
\end{figure}

Figure~\ref{fig:wespeaker} shows the overall structure and pipeline of Wespeaker. A standard procedure contains data preparation on the disk, online feature preparation, and model training. After the converged model is obtained, it can be easily exported to a run-time format and ready for further deployment. The extracted speaker embeddings can then be applied to downstream tasks, such as speaker verification and diarization.

\subsection{Data management}
\subsubsection{Unified IO}
\begin{figure}[!htb]
  \centering
  \vspace{0cm}
  \includegraphics[width=0.9\linewidth, trim=0 5 0 0, clip]{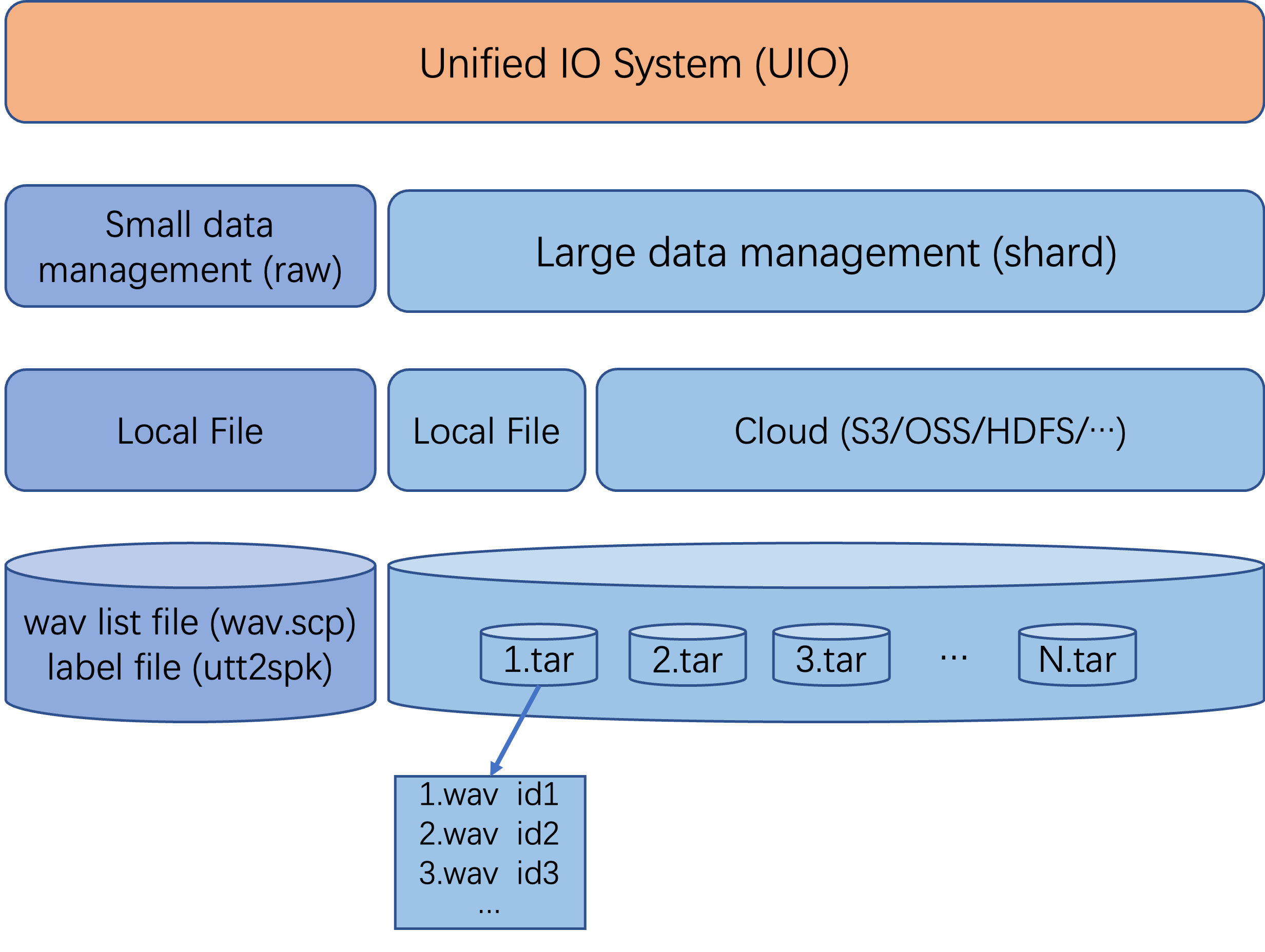}
  \caption{Unified IO in Wespeaker}
  \label{fig:uio}
  \vspace{-0.3cm}
\end{figure}

Production-scale corpus usually contains tens of thousands of hours of 
speech, which are comprised of massive small files. To avoid the possible consequent out-of-memory (OOM) and slow-training problems, we introduce the unified IO (UIO) mechanism in Wenet\footnote{\url{https://wenet.org.cn/wenet/UIO.html}} to the data management in Wespeaker. This mechanism was inspired by the TFRecord format used in Tensorflow~\cite{abadi2016tensorflow} and AIStore~\cite{aizman2019high}, which packs each set of small files into a bigger shard via the GNU tar.
As shown in Figure~\ref{fig:uio}, for the large dataset, on-the-fly decompression will be performed to sequentially read the shard files into the memory during the training stage. On the other hand, for the small dataset, Wespeaker supports the traditional data loading functions to load the raw files from the disk directly.

\subsubsection{On-the-fly feature preparation}
Traditional feature preparation for speaker embedding learning is usually done offline.
A typical offline procedure could comprise resampling, data augmentation, data slicing\footnote{Data slicing means cutting each utterance into a fixed length, which eases the data preparation and speeds up the GPU training. For text-independent speaker modeling, we assume this context information corruption has a limited impact on the modeling accuracy.} and feature extraction.
The offline feature preparation generates the final training examples and saves them on the disk, which will remain unchanged during the whole training process\footnote{We name this ``feat'' data type and also support it in Wespeaker.}.
Wespeaker loads the original wave data and performs all these steps in an on-the-fly manner, which has two main advantages: 1) There is no need to save the augmented wave files and processed features, which significantly saves the disk cost. 2) Online augmentation makes it possible for the model to see different training examples at different epochs, this uncertainty and randomness improve the robustness of the resultant model.

\begin{figure*}[!htb]
  \centering
  \vspace{0cm}
  \includegraphics[width=0.9\linewidth]{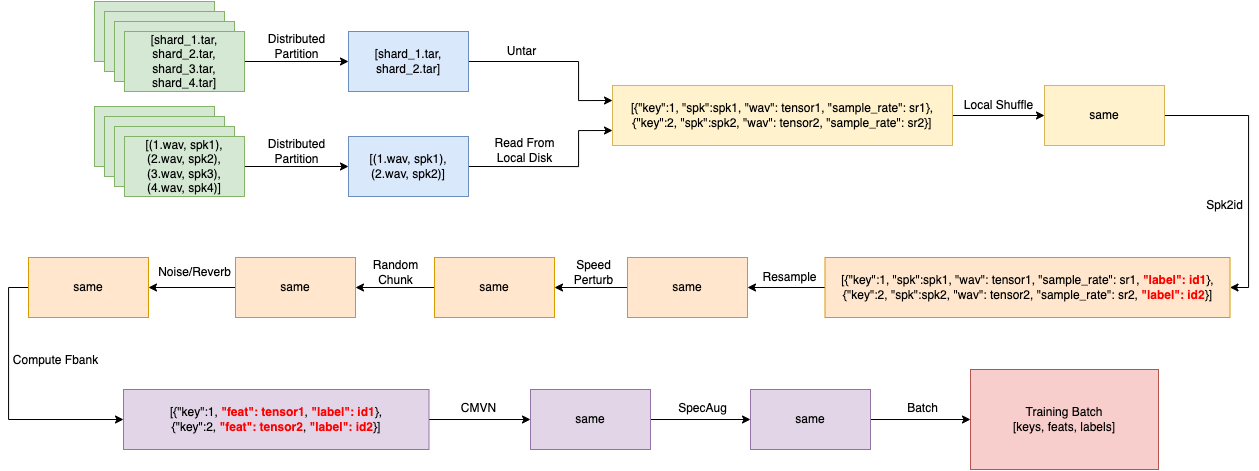}
  \caption{The pipeline of online feature preparation in Wespeaker. ``same'' means that no new attributes are added and only change a specific attribute value for each sample (except local shuffle), compared with the last node.}
  \label{fig:feat_pre}
  \vspace{-0.2cm}
\end{figure*}

Figure~\ref{fig:feat_pre} presents the pipeline of online feature preparation in Wespeaker, which includes the following modules:
\begin{itemize}
    \item Local shuffle: Construct a local buffer and shuffle it randomly each time, contributing to different training batches at different epochs.
    %It is recommended that the buffer size should be larger than the data number of a single shard. 
    \item Spk2id: Map speaker name into speaker id (from 0 to $N$-1, where $N$ is the total speaker number of the training set).
    \item Resample: Resample the training data into the targeted sample rate.
    \item Speed perturb: Change the speed of the training data with a certain probability.
    \item Random chunk: Chunk the training data into the same length randomly. Padding is applied on the short ones.
    \item Noise/Reverb: Add noise or reverberation augmentation.
    \item Compute Fbank: Extract Fbank feature from raw PCM data.
    \item CMVN: Apply cepstral mean and variance normalization per utterance.
    \item SpecAug\footnote{SpecAug is commonly not compatible with other augmentation types and thus set to False by default.}: Apply Spec-Augmentation~\cite{park2019specaugment} on the feature.
    \item Batch: Organize the training data into fixed-size batches.
\end{itemize}

\subsection{SOTA Model Implementation}
Wespeaker currently supports the following models, 
\begin{itemize}
    \item TDNN based x-vector, this is a milestone work that leads the following deep speaker embedding era.
    \item ResNet based r-vector and its deeper version, this is the winning system of VoxSRC 2019~\cite{zeinali2019but} and CNSRC 2022~\cite{chen2022sjtu}.
    \item ECAPA-TDNN, a modified version of TDNN, this is the winning system of VoxSRC 2020~\cite{desplanques2020ecapa}.
\end{itemize}

\textbf{Pooling functions} aggregate frame-level features into segment-level representations, where Wespeaker supports the statistics-based and attention-based ones.
\textbf{Loss functions} are critical to current deep speaker embedding learning. We support the standard softmax cross-entropy loss and different margin-based variants~\cite{hajibabaei2018unified,xiang2019margin}, such as A-softmax~\cite{liu2017sphereface,huang2018angular}, AM-softmax~\cite{wang2018additive} and AAM-softmax~\cite{deng2018arcface}.

\subsection{Training strategies}
\subsubsection{Learning Rate and Margin Schedule}
In Wespeaker implementation, the learning rate schedule is the result of two functions working together. Here, we denote the variation function of the learning rate with respect to time as $lr(t)$ and it can be represented by the product of warmup function $g(t)$ and exponential descent function $h(t)$: $lr(t)=g(t)h(t)$. The specific expressions of $g(t)$ and $h(t)$ are:

\begin{equation}
  g(t)=\begin{cases}
    \frac{t}{T_{warm}}, & \text{$t < T_{warm}$}.\\
    1, & \text{$T_{warm} \leq t<T$}.
  \end{cases}
\end{equation}
\begin{equation}
  h(t)= \eta_{0}\cdot e^{\left(\frac{t}{T} ln (\frac{\eta_{T}}{\eta_{0}})\right)}
\end{equation}
where $t$ is the current training iteration, $T_{warm}$ is the warm-up iteration, $T$ is the total iteration, $\eta_{0}$ is the initial learning rate and $\eta_{T}$ is the final learning rate.

The scheduler for the margin in loss is a three-stage function $m(t)$:

\begin{equation}
  m(t)=\begin{cases}
    0, & \text{$t<T_1$}.\\
    f(t), & \text{$T_1\leq t<T_2$}. \\
    M, & \text{$T_2\leq t<T$}.
  \end{cases}
\end{equation}
where $0 \leq T_1 \leq T_2 \leq T$, $M$ is the final margin in loss and $f(t)$ is a linear growth function or a logarithmic growth function from 0 to $M$.

\subsubsection{Large Margin Fine-tuning}
The large margin fine-tuning strategy was first proposed in~\cite{thienpondt2021idlab} and widely used in speaker verification challenge systems~\cite{chen2022sjtu, zhao2021speakin, makarov2022id} to further enhance the system's performance. This strategy is performed as an additional fine-tuning stage based on a well-trained speaker verification model. In this stage, the model will be trained with a larger margin and longer training segments relative to the normal training stage. For Wespeaker implementation, we use AAM loss with a margin of 0.5 and 6s training segments.

% \begin{itemize}
%     \item Learning rate scheduling
%     \item Margin scheduling
%     \item Large margin fine-tuning
% \end{itemize}

\subsection{Back-end Support}
For the dominant deep speaker embeddings supervised by large-margin softmax losses, the simple cosine similarity can serve as a good scoring back-end. Before the era of large-margin embeddings, parametric back-ends such as probabilistic linear discriminant analysis (PLDA) are more widely used. Wespeaker implements both scoring back-ends\footnote{The two-covariance version of PLDA is supported currently, more variants will be added in the future}, while an additional score normalization function~\cite{matejka2017analysis} is also provided to calibrate the speaker verification scores.

\subsection{Deployment}
For the models trained in Wespeaker, we can easily export them to ``tensorrt'' or ``onnx'' format, which can be deployed on the Triton Inference Server. Detailed information including the instructions and performance can be found at \url{https://github.com/wenet-e2e/wespeaker/tree/master/runtime/server/x86_gpu}. Furthermore, since Wespeaker is designed as a general speaker embedding learner, we also provide the python bindings and deploy it via standard ``pip'' packaging, which allows users to trivially use the pre-trained models for down-stream tasks\footnote{\url{https://github.com/wenet-e2e/wespeaker/tree/master/runtime/binding/python}, a toy demo on speaker verification can be found at \url{https://huggingface.co/spaces/wenet/wespeaker_demo}}.

\section{Experiments and Recipes}
As described in the above sections, deep speaker embeddings could be applied to different downstream tasks, whereas this paper focuses on speaker verification and speaker diarization.

\subsection{Basic setups for speaker embedding learning}
For the training setups of all speaker models in the following sections, we adopted the shard UIO method and applied the same online data augmentation in the training process.
%\begin{itemize}
%    \item Additive Noise: Audios from the MUSAN dataset~\cite{musan2015} are used as additive noises.
%    \item Reverberation: The simulated room impulse responses (RIRs)~\cite{ko2017study} are used for the reverberation.
%    \item Speed Perturbation: We randomly change the speed of an utterance with ratio 0.9 or 1.1, and the augmented audios will be treated as from new speakers due to the pitch shift after the augmentation. 
%\end{itemize}
%All above three augmentations are done in an on-the-fly manner.
The audios from the MUSAN dataset~\cite{musan2015} are used as additive noises, while the simulated room impulse responses (RIRs)\footnote{\url{https://www.openslr.org/28}} are used for the reverberation.
For each utterance in the training set, we apply additive-noise or reverberation augmentation (not both at the same time) with a probability of 0.6.
For speed perturbation, we randomly change the speed of an utterance with a ratio of 0.9 or 1.1, and the augmented audios will be treated as from new speakers due to the pitch shift after the augmentation.
Moreover, the ratio of speeds 0.9, 1.0 and 1.1 is set as 1:1:1.
The acoustic features are 80-dimensional log Mel-filter banks (Fbank) with a 10ms frameshift and a 25ms frame window.
All training data are chunked into 200 frames and CMN (without CVN) is also applied.
Note that SpecAug is not used in all experiments.

\subsection{Speaker Verification}
For the speaker verification task, recipes are constructed based on the VoxCeleb and CNCeleb datasets, which are very popular and used by many researchers, thanks to the promotion by related competitions. All the results exhibited here are obtained after large margin fine-tuning, with cosine scoring and AS-Norm applied.\footnote{Due to the limit of paper length, we only list the main results in the following experiments. More detailed results including ablation study, PLDA results, etc. can be found in the corresponding recipes online.}.
\subsubsection{VoxCeleb}
VoxCeleb dataset~\cite{nagrani2020voxceleb} was released by Oxford and has become one of the most popular text-independent speaker recognition datasets. Following the segmentation of the VoxSRC challenge, we only use the VoxCeleb2 dev as the training set, which contains more than one million audios from 5994 speakers.

\begin{table}[!htb]
\footnotesize
\centering
\vspace{-0.3cm}
\caption{\label{tab:vox_basic}Results achieved using different architectures on the VoxCeleb dataset, ``dev'' of part 2 is used as the training set}
\begin{adjustbox}{width=.45\textwidth,center}
\begin{tabular}{ccccccc}
\toprule
\multirow{2}{*}{Architecture} & \multicolumn{2}{c}{voxceleb1\_O} & \multicolumn{2}{c}{voxceleb1\_E} & \multicolumn{2}{c}{voxceleb1\_H}  \\
& EER(\%) & minDCF & EER(\%) & minDCF & EER(\%) & minDCF \\\midrule
TDNN & 1.590 & 0.166 & 1.641 & 0.170 & 2.726 & 0.248\\ \midrule
ECAPA-TDNN (\cite{desplanques2020ecapa})& 0.870 & 0.107 & 1.120 & 0.132 & 2.120 & 0.210 \\\midrule
ECAPA-TDNN & 0.728 & 0.099 & 0.929 & 0.100 & 1.721 & 0.169\\ \midrule
ResNet34 (\cite{zeinali2019but}) & 1.31 & 0.154 & 1.38 & 0.163 & 2.50 & 0.233\\ \midrule
ResNet34 & 0.723 & 0.069 & 0.867 & 0.097 & 1.532 & 0.146\\ \midrule
ResNet221 & 0.505 & 0.045 & 0.676 & 0.067 & 1.213 & 0.111\\ \midrule
ResNet293 & \textbf{0.447} & \textbf{0.043} & \textbf{0.657} & \textbf{0.066} & \textbf{1.183} & \textbf{0.111} \\ 
\bottomrule
\end{tabular}
\end{adjustbox}
\vspace{-0.1cm}
\end{table}

The results in Table~\ref{tab:vox_basic} show that our implementation achieves very competitive numbers compared with the original ones in the literature. Scaling the ResNet deeper can further boost the performance significantly.
% \begin{table}[!htb]
% \footnotesize
% \centering

% \caption{\label{tab:vox_tricks} Impact of large margin fine-tuning}
% \begin{adjustbox}{width=.45\textwidth,center}
% \begin{tabular}{ccccccc}
% \toprule
% \multirow{2}{*}{Architecture} & \multicolumn{2}{c}{voxceleb1\_O} & \multicolumn{2}{c}{voxceleb1\_E} & \multicolumn{2}{c}{voxceleb1\_H}  \\
% & EER & minDCF & EER & minDCF & EER & minDCF \\\midrule
% ResNet34 (basic) & \\ \midrule
% + LM finetune & \\ \midrule\midrule
% ECAPA-TDNN  & \\ \midrule
% + LM finetune & \\ 
% \bottomrule
% \end{tabular}
% \end{adjustbox}
% \end{table}

\subsubsection{CNCeleb}
For the CNCeleb recipe, we combine the 1996 speakers from CNCeleb2 and 797 speakers from the CNCeleb1 dev set as the training set, and evaluate on the CNCeleb1 test set.
Although the collection procedure of the CNCeleb dataset~\cite{li2022cn} is similar to the one of VoxCeleb, many recordings are shorter than 2 seconds in this dataset.
Therefore, in the data preparation, we first concatenate the short audios from the same genre and same speaker to construct audios longer than 5 seconds.

\begin{table}[!htb]
\footnotesize
\centering
\vspace{-0.3cm}
\caption{\label{tab:cn_results} Results on the CNCeleb evaluation set}
\begin{adjustbox}{width=.25\textwidth,center}
\begin{tabular}{ccc}
\toprule
Architecture & EER(\%) & minDCF  \\\midrule
TDNN & 8.960 & 0.446  \\ \midrule
ECAPA-TDNN  & 7.395 & 0.372 \\ \midrule
ResNet34(~\cite{tong2021asv}) & 9.141 & 0.463 \\ \midrule
ResNet34 & 6.492 & 0.354 \\ \midrule
ResNet221 & \textbf{5.655} & \textbf{0.330}\\ \bottomrule
\end{tabular}
\end{adjustbox}
\vspace{-0.1cm}
\end{table}

The results obtained using different backbones are exhibited in Table~\ref{tab:cn_results}.
Unlike the VoxCeleb evaluation protocol, CNCeleb assumes each speaker is enrolled with multiple sessions.
Embeddings for all enrollment sessions for each speaker are extracted and averaged to obtain the final enrollment embedding, which brings considerable performance improvement in our experiments.

\subsection{Speaker Diarization}
VoxConverse dataset~\cite{chung2020spot} was released for the diarization track in VoxSRC 2020, which is a ``in the wild'' dataset collected from the Youtube Videos. This recipe shows how to leverage a pre-trained speaker model for the speaker diarization task. The pre-trained ResNet34 model is used for speaker embedding extraction and spectral clustering is implemented specifically for this task.
As illustrated in Table~\ref{tab:voxconverse_results}, we achieve strong results on the VoxConverse dev dataset, using oracle speech activity detection (SAD) annotations or system SAD results from Silero-VAD~\cite{SileroVAD} pre-trained model, proving the effectiveness of deep speaker embedding learning in Wespeaker.

\begin{table}[!htb]
\footnotesize
\setlength\tabcolsep{3pt}
\centering
\vspace{-0.3cm}
\caption{\label{tab:voxconverse_results} Results on the VoxConverse dev set}
\begin{tabular}{ccccc}
\toprule
System & MISS(\%) & FA(\%) & SC(\%) & DER(\%)  \\ \midrule
\cite{chung2020spot} (system SAD) & 2.4 & 2.3 & 3.0 & 7.7 \\ \midrule
Wespeaker (system SAD)  & 4.4 & 0.6 & 2.1 & 7.1 \\ \midrule
Wespeaker (oracle SAD) &  2.3 &	0.0	& 1.9 &	4.2 \\ \bottomrule
\end{tabular}
\vspace{-0.2cm}
\end{table}

\section{Conclusion and future work}
In this paper, we introduced Wespeaker, a research and production oriented speaker embedding learning toolkit. Wespeaker has a lightweight code base and focuses on high-quality speaker embedding learning, achieving very competitive results on several datasets. Despite the friendliness for researchers, CPU-and GPU- compatible deployment codes are also integrated to bridge the gap between the research and production systems.

For the next release, we will focus on the following key points: 1) Self-supervised learning (SSL) for speaker embedding learning. 2) Small footprint solutions for resource-limited scenarios. 3) Continually adding SOTA speaker models (network architectures and scoring back-ends) and optimizing the training strategies.

% \section{Acknowledgement}
% \newpage
\footnotesize
\bibliographystyle{IEEEbib}
\bibliography{strings,refs}

\end{document}